\documentclass[notitlepage,english,aps,floats,onecolumn,showpacs,nofootinbib,floatfix]{revtex4-1}

\usepackage{pslatex}
\usepackage[T1]{fontenc}
\usepackage[latin1]{inputenc}
\usepackage{graphicx}
\usepackage{epsfig}
\usepackage{longtable}
\usepackage{float}
\usepackage{calc}
\usepackage{ifthen}
\usepackage{amsmath}
\usepackage{hyperref}
\usepackage{amssymb}

\usepackage{color}

{
{
{
\newcommand{\bea}{\begin{eqnarray}}
\newcommand{\eea}{\end{eqnarray}}

\newcommand{\nc}{\newcommand}
\nc{\renc}{\renewcommand}
\nc{\eqs}[2]{\mbox{Eqs.~(\ref{#1},\,\ref{#2})}}
\nc{\eq}[1]{\mbox{Eq.~(\ref{#1})}}
\nc{\figs}[2]{\mbox{Figs.~(\ref{#1},\,\ref{#2})}}
\nc{\fig}[1]{\mbox{Fig~.(\ref{#1})}}
\nc{\be}[1]{\begin{equation} \mbox{$\label{#1}$}}
\nc{\ee}{\vspace{0.1cm}\end{equation}}

\newcommand{\bean}{\begin{eqnarray*}}
\newcommand{\eean}{\end{eqnarray*}}

\def\GeV{{\rm \ GeV}}

\begin{document}

\title{The KSVZ Axion Model with Quasi-Degenerate Minima: A Unified Model for Dark Matter and Dark Energy} 

\author{Amy Lloyd-Stubbs  and John McDonald }
\email{a.lloyd-stubbs@lancaster.ac.uk}
\email{j.mcdonald@lancaster.ac.uk}
\affiliation{Dept. of Physics,  
Lancaster University, Lancaster LA1 4YB, UK}

\begin{abstract}

  We consider the possibility that dark matter and dark energy can be explained by the minimal KSVZ axion model. This is possible if the lowest energy minimum of the scalar potential has zero energy density, as can occur in theoretical models of vacuum energy cancellation based on spacetime averaging and in models based on energy parity.  
Dark energy is then understood as being due to the energy density of the metastable electroweak vacuum relative to a second quasi-degenerate minimum. The requirement of  quasi-degenerate minima is a non-trivial condition which completely determines the form of the potential for a given value of the axion decay constant, $f_{a}$, and the PQ scalar self-coupling, $\lambda_{\phi}$. The existence of the second quasi-degenerate minimum imposes a new lower bound on the axion decay constant, $f_{a} \geq 2.39 \times 10^{10} \, \lambda_{\phi}^{-1/4} \GeV$. If the PQ symmetry is broken after inflation then the lower bound on $f_{a}$ implies a lower bound on the amount of axion dark matter, $\Omega_{a}/\Omega_{dm} \geq (0.28-0.46)\,\lambda_{\phi}^{-0.291}$, where the range is due to the uncertainty in the amount of axion dark matter produced by vacuum realignment, cosmic strings and domain walls. Therefore at least 30$\%$ of dark matter must be due to axions if $\lambda_{\phi} \lesssim 1$.
If axions constitute all of the dark matter then the value of $f_{a}$, and so the form of the scalar potential, is completely fixed for a given value of $\lambda_{\phi}$, with only a weak dependence on $\lambda_{\phi}$. This will allow the inflation and post-inflation evolution of the model to be quantitatively studied for a given inflation model and dimensionally natural values of $\lambda_{\phi}$.  

\end{abstract}
 
\maketitle

\section{Introduction}

The Standard Model (SM) has been successful in explaining the observed nature of strong and electroweak interactions. However, there are fundamental questions which are not addressed by the SM. Here we will consider an existing extension of the SM, the KSVZ version of the Pecci-Quinn (PQ) axion model \cite{ksvz, pq}, which is presently able to address two of these questions, the strong CP problem and the nature of dark matter. We will show that in the case where the energy density is zero at the lowest minimum of the scalar potential, the KSVZ axion model can also account for dark energy in the form of the energy density of the electroweak vacuum state relative to a second minimum. We will also show that the requirement that the scalar potential is very close to the limit of degenerate vacuum states is a non-trivial constraint.  This has implications for the minimum amount of axion dark matter and for the form of the scalar potential for a given axion decay constant. 

  One class of models which can account for the lowest energy minimum being at zero energy density are those based on spacetime averaging, where the energy averaged over spacetime adds a negative constant energy density to the Lagrangian. If the universe is in the lowest energy vacuum state over most of the volume of spacetime, then the energy of that vacuum state will be cancelled. The first model of this type was the universe multiplication model proposed in \cite{linde}. Other more recent approaches include vacuum sequestering \cite{padilla} and a model based on non-local constraints \cite{carroll}.  A different approach, which can also result in the lowest energy minimum having zero energy, is based on energy parity \cite{kaplan}. In this case the energy density of the lowest energy minimum is zero if the mirror universe occupies the same vacuum state. In all of these models, all of which seek to explain the absence of a large vacuum energy density, the total energy of one of the vacuum states is effectively subtracted away. This also means that there can be no cosmological constant. Therefore the observed dark energy density must in be explained by the potential energy of a scalar field.

The KSVZ axion model adds a singlet complex scalar field, the PQ scalar, $\Phi$, to the scalar sector of the SM. The axion is the angular mode of $\Phi$. The axial anomaly couples the axion to the QCD topological term $G\tilde{G}$, which generates a potential for the axion after the QCD phase transition. The axion field cancels the CP-violating QCD $\theta$-angle at the minimum of this potential, solving the strong-CP problem. 

   Extending the SM by the addition of $\Phi$ modifies the classical scalar potential of the model to a function of two scalar fields, $h$ and $\phi$, where $h$ is the electroweak doublet Higgs scalar and $\phi$ is the radial mode of the PQ scalar. The minimum of the classical potential for a given value of $h$ is then a curving trajectory in the $(h, \phi)$ plane. Therefore the question of the stability of the electroweak vacuum is quite different from the case of the SM. In addition, the observed SM Higgs boson is a mixture of $\phi$ and $h$. As a result, the electroweak doublet Higgs quartic coupling can be larger than the SM Higgs quartic coupling which is determined experimentally from the Higgs mass. The effect of increasing the electroweak Higgs doublet quartic coupling is to increase the stability of the electroweak vacuum with respect to quantum corrections. The combined effect of the classical modification of the scalar potential and the increase of the electroweak Higgs doublet quartic coupling was first discussed in \cite{threshold}, where it was called the scalar threshold effect. 

    In this paper we will show how the combination of the classical scalar potential and quantum corrections can allow the existence of a second vacuum, which is at a slightly lower energy than the electroweak vacuum. In the case where there is zero energy density at the lowest energy minimum of the potential, the observed dark energy density can then be explained by the vacuum energy of the metastable electroweak minimum relative to the lowest energy minimum. (The possibility of metastable dark energy has previously been considered in \cite{meta1,meta2,meta3,krauss,landim,landim2}.) Combined with axion dark matter, the KSVZ axion model can then serve as a minimal model for both dark matter and dark energy, as well as providing a solution to the strong CP problem.

   In the case where the PQ symmetry is restored after inflation and subsequently broken, as may be necessary in order to satisfy axion isocurvature constraints, the axion dark matter density is determined for a given value of $f_{a}$. We will show that the requirement of a quasi-degenerate second minimum imposes a lower bound on $\lambda^{1/4}_{\phi} f_{a}$, where $\lambda_{\phi}$ is the PQ scalar self-coupling. This in turn imposes a lower bound on the density of axion dark matter. 

In addition, we will show that the form of the potential is entirely determined by the value of $\lambda_{\phi}^{1/4}f_{a}$. Therefore, in the case of PQ symmetry breaking after inflation,  the potential is determined by the axion dark matter density for a given value of $\lambda_{\phi}$, with only a weak dependence on this unknown coupling.  

   Our paper is organized as follows. In Section 2 we review the KSVZ axion model and explain our method for determining the quantum corrections to its potential. 
In Section 3 we discuss the conditions for a second quasi-degenerate vacuum as a function of the axion decay constant. In Section 4 we discuss the cosmology of the quasi-degenerate potential, in particular for the case where the PQ symmetry is broken after inflation.  In Section 5 we present our conclusions.

\section{The scalar potential of the KSVZ axion model}

\subsection{The Classical Scalar Potential} 

We will consider the minimal KSVZ axion model in the following, which adds a singlet complex scalar field to the scalar sector of the SM\footnote{For a previous study of the KSVZ axion model with the scalar threshold effect, see \cite{salvio}.}. The Lagrangian fermion term added to the SM is
\be{l1} {\cal L}_{Q} =  y_{Q}  \overline{Q}_{R} Q_{L} \Phi \;\;\; + \; {\rm h.c.}     ~.\ee
Here $Q$ is a triplet of fermions under $SU(3)_{c}$ which is a singlet under the electroweak group, and $\Phi = \phi/\sqrt{2} e^{i a/f_{a}}$ is the PQ scalar, where $\phi$ is the radial PQ scalar and $a$ is the axion. These terms are invariant under an axial global 
$U(1)_{PQ}$ symmetry where $\Phi \rightarrow e^{i \alpha} \Phi$ and $Q \rightarrow e^{i \alpha \gamma_{5}} Q$. The axial
anomaly of the $U(1)_{PQ}$ symmetry couples the axion to the 
QCD topological term, which generates the necessary potential 
to solve the strong CP problem\footnote{In the case where there is only one triplet, $Q$, the domain wall number $N_{DW}$ is equal to 1. In this case there is no axion domain wall problem.}.

The scalar potential of the model is given by 
\be{v1}   V(H, \Phi) = \lambda_{h} \left(|H|^2 - \frac{v^2}{2} \right)^2 + \lambda_{\phi}\left(|\Phi|^2 - \frac{f_{a}^{2}}{2} \right)^2 + \lambda_{h \phi} \left(|H|^2 - \frac{v^2}{2} \right) \left( |\Phi|^2 - \frac{f_{a}^{2}}{2} \right)   ~.\ee
where $H$ is the Higgs doublet.
As a function of the electroweak doublet Higgs field $h$ and radial PQ scalar $\phi$, this becomes 
\be{v2}   V(h, \phi) = \frac{\lambda_{h}}{4} \left(h^2 - v^2 \right)^2 + \frac{\lambda_{\phi}}{4}\left(\phi^2 - f_{a}^{2} \right)^2 + \frac{\lambda_{h \phi}}{4} \left(h^2 - v^2\right) \left(\phi^2 - f_{a}^{2}\right)   ~.\ee
To determine the minima of this potential, we will consider the minimum in the $\phi$ direction as a function of $h$, $\phi(h)$, 
then follow the trajectory of this minimum in the $(h, \phi)$ plane. $\phi(h)$ is given by 
\be{v3}  \phi^{2}(h) = \frac{2}{\lambda_{\phi}} \left( \frac{\mu_{\Phi}^{2}}{2} - \frac{\lambda_{h \phi}}{4} h^2 \right)  ~,\ee
where 
\be{v9} \mu_{\Phi}^{2} = \lambda_{\phi} f_{a}^{2} + \frac{\lambda_{h\phi}}{2} v^2    ~.\ee
This is true when $h < h_{c}$, where $h_{c}$ is the value at which $\phi^{2}(h) = 0$ and the $U(1)_{PQ}$ symmetry is restored,
\be{v4} h_{c} = \sqrt{ \frac{2 \mu_{\Phi}^{2} }{\lambda_{h\phi}} }    ~.\ee
In this we have assumed that $\lambda_{h \phi} > 0$. If $\lambda_{h \phi} < 0$ then the $U(1)_{PQ}$ symmetry is not restored as $h$ increases, in which case there can be no degenerate minima.   Since $\mu_{\Phi}^{2} \approx \lambda_{\phi} f_{a}^{2}$ to a good approximation for natural values of the couplings,  we can also write \eq{v4} as 
\be{v4a} h_{c} =    \left(\frac{4 \lambda_{\phi}}{\lambda_{h \phi}^{2}} \right)^{1/4} \times \lambda_{\phi}^{1/4}  f_{a} ~.\ee    
This will be important in the following. The potential along the minimum trajectory as a function of $h$ when $h < h_{c}$  is then 
\be{v4b} V(h, \phi(h)) = - \left( \lambda_{h} - \frac{\lambda_{h\phi}^{2}}{4 \lambda_{\phi}} \right) \frac{v^2}{2} h^2 + \frac{1}{4} \left(\lambda_{h} - \frac{\lambda_{h\phi}^{2}}{4 \lambda_{\phi}} \right) h^4 +\frac{1}{4} \left( \lambda_{h} -  \frac{\lambda_{h\phi}^{2}}{4 \lambda_{\phi}} \right) v^4     ~.\ee
At $h > h_{c}$, $\phi(h) = 0$ and so the potential along the minimum trajectory becomes 
\be{v5}   V(h, 0) =  -\frac{\mu_{H}^{2}}{2} h^2 + \frac{\lambda_{h}}{4} h^4 + \Lambda    ~,\ee
where 
\be{v6} \mu_{H}^{2} = \lambda_{h} v^{2} + \frac{\lambda_{h\phi}}{2} f_{a}^2    ~\ee
and 
\be{v7}   \Lambda = \frac{\lambda_{h}}{4} v^4 + \frac{\lambda_{\phi}}{4} f_{a}^{4} + \frac{\lambda_{h\phi}}{4} v^2 f_{a}^{2}    ~.\ee
The potential along the minimum trajectory depends purely on the combination $\lambda_{h\phi}^{2}/\lambda_{\phi}$. The only explicit dependence on the PQ scalar couplings $\lambda_{h \phi}$ and $\lambda_{\phi}$ in our analysis is in the relation between $h_{c}$ and $f_{a}$ in \eq{v4a}, where there is a weak $\lambda_{\phi}^{1/4}$ dependence. 

A key feature of the potential along the minimum as function of $h$ is the change in behaviour of the $h^4$ term at $h = h_{c}$, where the effective quartic coupling changes from $\lambda_{h} - \lambda_{h \phi}^{2}/4 \lambda_{\phi}$ at $h < h_{c}$ to $\lambda_{h}$ at $h > h_{c}$. This jump in the value of the effective quartic coupling will play an important role in producing the degenerate minimum of the potential at $h$ close to $h_{c}$ once quantum corrections are included.

In addition to creating a two-field potential, the addition of the PQ scalar also modifies the input value of the Higgs doublet quartic coupling $\lambda_{h}$ at the electroweak scale. This is due to mixing between $h$ and $\phi$. On expanding the field around the electroweak vacuum at $(h, \phi) = (v, f_{a})$ as $h = \hat{h} + v$ and $\phi = \hat{\phi} + f_{a}$, the mass matrix of $\hat{h}$ and $\hat{\phi}$ becomes
\be{v8}   \begin{pmatrix} \frac{\lambda_{h\phi}}{2} f_{a}^2 + 3 \lambda_{h} v^2 - \mu_{H}^{2} & \lambda_{h\phi} v f_{a} \\
\lambda_{h\phi} v f_{a} & \frac{\lambda_{h\phi}}{2} v^2 + 3 \lambda_{\phi} f_{a}^2 - \mu_{\Phi}^{2}  \end{pmatrix}   ~.\ee
Using $f_{a}^2 \gg v f_{a} \gg v^2$, the eigenvalues of this matrix are 
\be{v10}   m_{h_{SM}}^{2} \approx 2 \left(\lambda_{h} - \frac{\lambda_{h\phi}^{2}}{4 \lambda_{\phi}} \right)v^2    ~\ee
and
\be{v11}   m_{\phi'}^2 \approx 2 \lambda_{\phi} f_{a}^2     ~,\ee
where $h_{SM}$ is the observed SM Higgs boson with $m_{h_{SM}} = 125$ GeV and $\phi'$ is the orthogonal mass eigenstate. The mass eigenstates are related to $\hat{h}$ and $\hat{\phi}$ by 
\be{v12}  h_{SM} = \alpha \hat{h} - \beta \hat{\phi}  ~\ee
and 
\be{v12a} \phi' = \alpha \hat{\phi} + \beta \hat{h}  ~,\ee 
where
\be{v13}  \beta = \frac{\lambda_{h \phi}}{2 \lambda_{\phi}} \frac{v}{f_{a}}  \ll 1 \;\;\;;\;\;\; \alpha = \sqrt{1 - \beta^2} \approx 1 ~.\ee
Although the SM Higgs $h_{SM}$ has only a very small admixture of $\hat{\phi}$, the large value of $f_{a}$ relative to $v$ causes a significant shift of the relation between the SM Higgs mass and the Higgs doublet quartic coupling, $\lambda_{h}$. The SM Higgs quartic coupling, $\lambda_{h \, SM}$, which is defined by 
$m_{h_{SM}}^2 = 2 \lambda_{h \, SM} v^2$,  is related to $\lambda_{h}$ by
\be{v14}     \lambda_{h \, SM}=  \lambda_{h} - \frac{\lambda_{h\phi}^{2}}{4 \lambda_{\phi}}      ~.\ee 
Thus $\lambda_{h}$ is generally larger than $\lambda_{h\,SM}$. The effect of the two-field potential and the increase of $\lambda_{h}$ relative to $\lambda_{h \, SM}$ is equivalent to the scalar threshold effect discussed in \cite{threshold}, where the heavy PQ scalar is integrated out to form an effective single field Higgs potential. In our analysis we will work with the full two-field scalar potential, without integrating out the heavy PQ scalar. 

\subsection{Quantum Corrections to the Scalar Potential}  

   In order to calculate the quantum corrections to the potential, we need to calculate the effective action as a function of two scalar fields. For a single scalar field the effective action can be expressed as a functional Taylor series
\be{p1} \Gamma[\varphi] = \sum_{n} \frac{1}{n!} \int d^{4}x_{1} ... 
d^{4}x_{n} \Gamma^{(n)}(x_{1}, ... , x_{n}) 
\varphi(x_{1}) ...  \varphi(x_{n})  ~,\ee
where $\varphi$ is the classical field.  
The effective potential then corresponds to the sum of 1PI Green's functions with all external particles at zero momentum 
\be{p2} V(\varphi) = - \sum_{n} \frac{1}{n!} \Gamma^{(n)}(p_{i} = 0\,; \mu) \varphi^{n}     ~,\ee
where $\Gamma^{(n)}(p_{i}\,; \mu)$ are the 1PI Green's functions in momentum space, renormalized at a scale $\mu$ in a given renormalization scheme. On performing the sum over $n$, the effective potential to 1-loop order becomes the   
tree-level (i.e. classical) potential plus the 1-loop Coleman-Weinberg (CW) correction. 

In the case with two scalar fields $\varphi$ and $\chi$, the effective action is a straightforward generalization of the functional Taylor expansion to the case with two scalar fields. The corresponding effective potential is therefore of the form   
\be{p3} V(\varphi, \chi) = - \sum_{l,m} \frac{1}{l! m!} \Gamma^{(l, m)}(p_{i} = 0\,; \mu) \varphi^{l} \chi^{m} ~.\ee
The functions $\Gamma^{(l,m)}(p_{i}\; ; \mu)$ are 1PI Green's functions with $l$ external $\varphi$ fields and $m$ external $\chi$ fields. However, there is then a problem when using perturbation theory to calculate the effective potential in the case where one of the fields has a large value compared to the renormalization scale of the theory whilst the other field remains small. For example, if $\varphi$ is large compared to the initial renormalization scale, $\mu_{0}$, while $\chi \sim \mu_{0}$, then on summing over the terms with only the $\varphi$ fields we will get a contribution to the effective potential which is just the conventional single field CW correction. We therefore need to run $\mu$ up to $\mu \sim \varphi$ to keep this contribution to the effective potential perturbative. However, this will then cause the CW correction calculated from the sum over only the $\chi$ terms to have a large logarithm, $\ln(\chi/\mu)$. Therefore it is difficult  to use perturbation theory to calculate the effective potential at points $(\varphi, \chi)$ where the magnitudes of $\varphi$ and $\chi$ are very different.

Although there are multiscale renormalization group methods which attempt to address this problem \cite{einhorn,ford,bando,bando2} (and which are still being developed \cite{steele,iso,prok}),  in this first analysis of the potential we will neglect the quantum effects of the $\lambda_{h \phi}$ and $\lambda_{\phi}$ by setting them to zero in the 1-loop Green's function calculations. This will produce the correct potential if the classical modification of the potential due to the $\phi$ field and its mixing with the electroweak doublet Higgs field dominates the effect of the quantum corrections due to $\lambda_{h \phi}$ and $\lambda_{\phi}$. This will be true if these couplings are sufficiently small. 
In this case we can set the 1-loop contribution to $\Gamma^{(l,m)}$ in \eq{p3} to zero for $m \neq 0$ and for any internal $\phi$ lines, in the case where $\varphi$ is identified with $h$ and $\chi$ with $\phi$. The quantum corrections to the effective potential then reduce to the sum over $h$ in the effective potential \eq{p3}, in which case the quantum corrections at the point $(h, \phi)$ are simply the SM CW corrections calculated for the field $h$. This will allow us to use the existing two-loop SM Renormalization Group (RG) equations to calculate the effective potential in a well-defined limit. It is always possible choose both $\lambda_{h\phi}$ and $\lambda_{\phi}$ to be small enough that quantum corrections due to these couplings are  negligible, whilst still obtaining a degenerate vacuum purely due to the classical effects of $\lambda_{h\phi}$ and $\lambda_{\phi}$. This is because main effects of the axion sector scalars on the stability of the electroweak vacuum, namely the classical modification of the quartic electroweak doublet coupling $\lambda_{h}$ relative to the SM Higgs coupling $\lambda_{h\,SM}$ and the form of the potential along the minimum trajectory, $V(h, \phi(h))$, both depend only on the combination $\lambda_{h\phi}^{2}/\lambda_{\phi}$, which can be large even if $\lambda_{h \phi}$ and $\lambda_{\phi}$ are small.

We define the tree-level classical potential using couplings evaluated at the renormalization point $\mu = m_{t}$ in the $\overline{{\rm MS}}$ scheme to be  $V_{0}(h, \phi; \mu = m_{t})$. There will also be a quantum correction to this potential at the point $(h, \phi)$, which we denote by $\Delta V_{q}(h, \phi)$. In order to calculate the quantum correction to the potential,  we will start from the electroweak vacuum state, $(h, \phi) = (v, f_{a})$. We first consider the potential in the direction corresponding to the electroweak doublet Higgs scalar $h$. Having calculated the correction at $(h,f_{a})$, we next fix the corresponding value of $h$ and move in the $\phi$ direction to the point of interest on the minimum trajectory, $(h, \phi(h))$. (The trajectory is schematically illustrated in Figure 1.) Since we are neglecting the quantum corrections due to $\lambda_{h\phi}$ and $\lambda_{\phi}$, only the classical potential changes as we move from $(h, f_{a})$ to  $(h, \phi(h) )$. 
The result of this is that the quantum corrected potential at the point $(h, \phi(h))$ is given by 
\be{p2} V(h, \phi(h)) = V_{0}(h, \phi(h); \mu = m_{t})
+ \Delta V_{q}(h, f_{a})    ~. \ee
To compute $\Delta V_{q}(h, f_{a})$, we run the 2-loop SM RG equations in the $\overline{{\rm MS}}$ scheme from the electroweak vacuum (with $\mu = m_{t}$) to $\mu = h$ to obtain the couplings at this renormalization point. We then add the 1-loop CW correction in the $\overline{\rm{MS}}$ scheme, $\Delta V_{CW}$. 

The tree-level potential along $(h, f_{a})$ at $h \gg v$ is simply
\be{p5}   V_{0}(h, f_{a}; \mu) \approx \frac{\lambda_{h}(\mu)}{4} h^{4}      ~.\ee 
Thus 
\be{p5b}   \Delta V_{q}(h, f_{a}) = \frac{\lambda_{h}(\mu = h)}{4} h^{4} - \frac{\lambda_{h}(\mu = m_{t})}{4} h^{4}
+ \Delta V_{CW}(h ; \mu = h)    ~.\ee
Therefore, from \eq{p2} and \eq{p5b}, the quantum correction to the potential along the minimum trajectory $(h, \phi(h))$ 
is given by simply replacing the $\lambda_{h} h^{4}/4$ term by $\lambda_{h}(\mu = h) h^{4}/4$ and adding the 1-loop CW term. 
Thus, at $h < h_{c}$,  the quantum corrected potential is 
\be{p6} V(h, \phi(h)) = - \left( \lambda_{h} - \frac{\lambda_{h\phi}^{2}}{4 \lambda_{\phi}} \right) \frac{v^2}{2} h^2 + \frac{1}{4} \left(\lambda_{h}(\mu = h) - \frac{\lambda_{h\phi}^{2}}{4 \lambda_{\phi}} \right) h^4 +\frac{1}{4} \left( \lambda_{h} -  \frac{\lambda_{h\phi}^{2}}{4 \lambda_{\phi}} \right) v^4  + \Delta V_{CW}(h; \mu = h)   ~,\ee
where all couplings are evaluated at $\mu = m_{t}$ except for $\lambda_{h}(\mu = h)$. At $h > h_{c}$, where $\phi(h) = 0$, the potential along the minimum trajectory becomes 
\be{p7}   V(h, 0) =  -\frac{\mu_{H}^{2}}{2} h^2 + \frac{\lambda_{h}(\mu = h)}{4} h^4 + \Lambda   + \Delta V_{CW}(h; \mu = h)  ~.\ee

\begin{figure}[htbp]
\begin{center}
\includegraphics[width=0.5\textwidth, angle=-90]{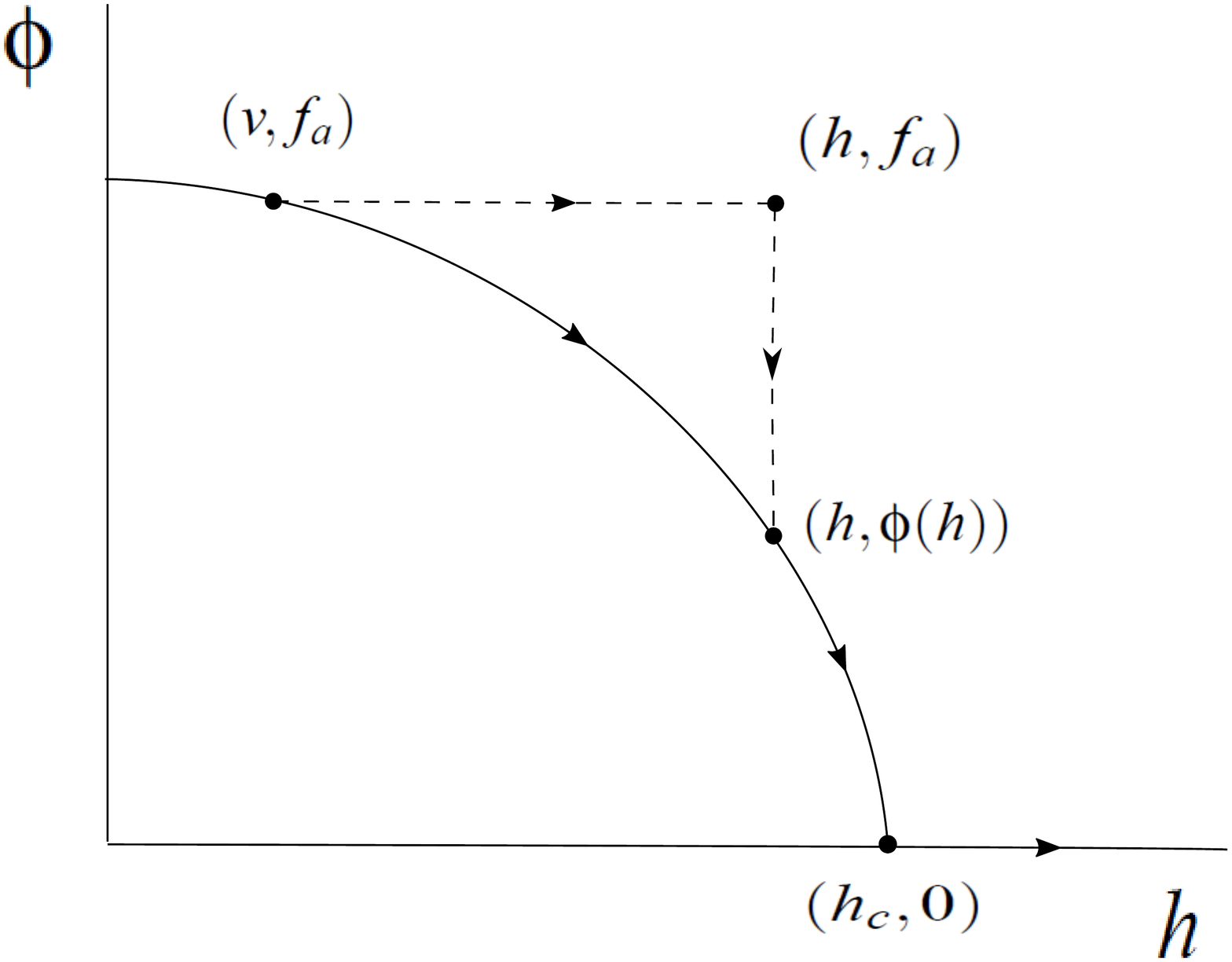} 
\caption{Illustration of the trajectory of the minimum of the potential in the $\phi$ direction as a function of $h$, $(h, \phi(h))$ (solid line), and the trajectory used to calculate the quantum correction to the potential (dashed line). } 
\label{fig1}
\end{center}
\end{figure}

\section{Quasi-Degenerate Minima of the Potential} 

To calculate the potential, we will use the 2-loop SM RG equations given in Appendix A of \cite{riotto} and the 1-loop SM CW potential in the $\overline{{\rm MS}}$ scheme given in \cite{drt}. 
We use the initial values for the SM couplings in the $\overline{{\rm MS}}$ scheme at $\mu = m_{t}$ as given in \cite{strumia}:  $\lambda_{h}(m_{t}) = 0.12604$, $y_{t}(m_{t}) = 0.93690$, $g_{2}(m_{t}) = 0.64779$, $g_{Y}(m_{t}) = 0.35830$, $g_{3}(m_{t}) = 1.6666$ and $m_{t} =173.34 \GeV$.

To explain dark energy, we require an effectively degenerate second minimum of the scalar potential at $V = 0$ in addition to the electroweak vacuum. (The tiny increase of the electroweak vacuum relative to the second minimum is negligible on the scale of the potential.) The potential is a function of a single parameter, which can be taken to be the combination $\lambda_{h\phi}/\sqrt{\lambda_{\phi}}$. To obtain the degenerate minimum, our procedure is to first consider the potential with $h_{c} >> h_{1}$, where $h_{1}$ is the value of $h$ at which $V = 0$ for a given value of $\lambda_{h\phi}/\sqrt{\lambda_{\phi}}$ when $h_{c} \gg h_{1}$. We next set $h_{c}$, the value of $h$ at which $\phi(h) = 0$, to a value close to $h_{1}$. The effect is to change the value of the quartic coupling from $\lambda_{h}(\mu) -\lambda_{h \phi}^{2}/4 \lambda_{\phi}$ at $h < h_{c}$ to $\lambda_{h}(\mu)$ at $h > h_{c}$ (where $\mu = h$). The jump in the quartic coupling causes the potential, which had been approaching zero at $ h < h_{c}$, to increase at $h > h_{c}$, hence generating a minimum. In practice we have to adjust $h_{c}$ to a value which is a little below $h_{1}$ in order to obtain a degenerate minimum with $V = 0$ once the RG running of $\lambda_{h}(\mu = h)$ is included. The value of $h_{c}$ then determines the value of $\lambda_{\phi}^{1/4} f_{a}$ via \eq{v4a}. 

In Table 1 we show\footnote{
The potential minima for the values of $\lambda_{h \phi}/\sqrt{\lambda_{\phi}}$ in Tables 1 and 2 are only approximately degenerate, to an accuracy determined by the degree of tuning of $\lambda_{\phi}^{1/4} f_{a}$ for a given value of $\lambda_{h \phi}/\sqrt{\lambda_{\phi}}$  and the precision of the code. As can be seen from Figures 1 to 4, the potentials are very close to degenerate relative to the scale of the potentials, with the deviation of the degenerate minimum at $h_{min}$ from zero being of the order of $10^{-3}$ of the potential at its maximum.} the values of $\lambda_{\phi}^{1/4} f_{a}$, $h_{min}$, $h_{max}$ and $V^{1/4}_{max}$ for the case of degenerate minima as a function of $\lambda_{h\phi}/\sqrt{\lambda_{\phi}}$. 
Here $h_{min}$ is the value of $h$ at the degenerate minimum, $h_{max}$ is the value of $h$ at the maximum of the potential between the electroweak vacuum and the second  vacuum along the minimum trajectory, and $V_{max}^{1/4}$ gives the value of the potential at the maximum.  We find that there is a finite range $\lambda_{\phi}^{1/4} f_{a}$ for which a degenerate minimum is possible, between 0.10 and 0.297. 
The upper bound corresponds to values of $h$ at the degenerate minimum which are greater than the Planck scale, which we assume is a natural cutoff for the effective theory.  
The lower bound corresponds to values of $\lambda_{h \phi}^{2}/ \lambda_{\phi}$ which are too small to sufficiently modify the potential to produce a second degenerate minimum. 
The lower bound on $\lambda_{h \phi}/\sqrt{\lambda_{\phi}}$ corresponds to a lower bound on $\lambda_{\phi}^{1/4} f_{a}$ equal to $2.39 \times 10^{10} \GeV$. This lower bound is an important prediction of metastable dark energy in the KSVZ axion model.

In Table 2 we give a more detailed set of parameters for the range of $\lambda_{\phi}^{1/4} f_{a}$ which is relevant to axion dark matter in the case where PQ symmetry is broken after inflation (discussed in the next section).

\begin{table}[h]
\begin{center}
\begin{tabular}{ |c|c|c|c|c|c| }
\hline
$\lambda_{h\phi}/\sqrt{\lambda_{\phi}}$ & $\lambda_{\phi}^{1/4} f_{a}$ &  $h_{max}/{\rm GeV}$ &$V_{max}^{1/4}/\GeV$ & $h_{min}/{\rm GeV}$  \\
\hline
$0.10$ & $2.39 \times 10^{10}$ & $8.89 \times 10^{10}$ & $9.50 \times 10^{9}$ & $1.26 \times 10^{11}$ \\
\hline
$0.12$ & $3.81 \times 10^{10}$ & $1.26 \times 10^{11}$ & $1.33 \times 10^{10}$ & $1.70 \times 10^{11}$ \\
\hline
$0.14$ & $6.49 \times 10^{10}$ & $1.96 \times 10^{11}$ & $2.05 \times 10^{10}$ & $2.59 \times 10^{11}$ \\
\hline
$0.16$ & $1.22 \times 10^{11}$ & $3.41 \times 10^{11}$ & $3.52 \times 10^{10}$ & $4.48 \times 10^{11}$ \\
\hline
$0.18$ & $2.61 \times 10^{11}$ & $6.85 \times 10^{11}$ & $6.94 \times 10^{10}$ & $8.96 \times 10^{11}$ \\
\hline
$0.20$ & $6.71 \times 10^{11}$ & $1.66 \times 10^{12}$ & $1.64 \times 10^{11}$ & $2.16 \times 10^{12}$ \\
\hline
$0.22$ & $2.23 \times 10^{12}$ & $5.26 \times 10^{12}$ & $5.04 \times 10^{11}$ & $6.80 \times 10^{12}$ \\
\hline
$0.24$ & $1.09 \times 10^{13}$ & $2.46 \times 10^{13}$ & $2.26 \times 10^{12}$ & $3.20 \times 10^{13}$ \\
\hline
$0.26$ & $1.04 \times 10^{14}$ & $2.24 \times 10^{14}$ & $1.93 \times 10^{13}$ & $2.90 \times 10^{14}$ \\
\hline
$0.28$ & $3.94 \times 10^{15}$ & $8.19 \times 10^{15}$ & $6.33 \times 10^{14}$ & $1.06 \times 10^{16}$ \\
\hline
$0.29$ & $9.31 \times 10^{16}$ & $1.90 \times 10^{17}$ & $1.32 \times 10^{16}$ & $2.45 \times 10^{17}$ \\
\hline
$0.297$ & $7.04 \times 10^{18}$ & $1.41 \times 10^{19}$ & $7.95 \times 10^{17}$ & $1.83 \times 10^{19}$ \\
\hline
\end{tabular}
\caption{Values of $\lambda_{h \phi}/\sqrt{\lambda_{\phi}}$ and the corresponding values of $\lambda_{\phi}^{1/4} f_{a}$ and $h_{min}$ for which the second quasi-degenerate minimum of the potential is obtained. We also show the values $h_{max}$ and $V^{1/4}_{max}$ at the maximum of the potential along the minimum trajectory.}
\end{center}
\end{table}

\begin{table}[h]
\begin{center}
\begin{tabular}{ |c|c|c|c|c|c| }
\hline
$\lambda_{h\phi}/\sqrt{\lambda_{\phi}}$ & $\lambda_{\phi}^{1/4} f_{a}$ &  $h_{max}/{\rm GeV}$ &$V_{max}^{1/4}/\GeV$ & $h_{min}/{\rm GeV}$  \\
\hline
$0.10$ & $2.39 \times 10^{10}$ & $8.89 \times 10^{10}$ & $9.50 \times 10^{9}$ & $1.26 \times 10^{11}$ \\
\hline
$0.105$ & $2.68 \times 10^{10}$ & $9.62 \times 10^{10}$ & $1.02 \times 10^{10}$ & $1.34 \times 10^{11}$ \\
\hline
$0.11$ & $3.00\times 10^{10}$ & $1.05 \times 10^{11}$ & $1.11 \times 10^{10}$ & $1.44 \times 10^{11}$ \\
\hline
$0.115$ & $3.38 \times 10^{10}$  & $1.15 \times 10^{11}$ & $1.216 \times 10^{10}$ & $1.56 \times 10^{11}$ \\
\hline
$0.12$ & $3.81 \times 10^{10}$ & $1.26 \times 10^{11}$ & $1.33 \times 10^{10}$ & $1.70 \times 10^{11}$ \\
\hline
$0.125$ & $4.32 \times 10^{10}$ & $1.39 \times 10^{11}$ & $1.47 \times 10^{10}$ & $1.89 \times 10^{11}$ \\
\hline
$0.13$ & $4.92\times 10^{10}$ & $1.55 \times 10^{11}$ & $1.63 \times 10^{10}$ & $2.07\times 10^{11}$ \\
\hline
$0.135$ & $5.63 \times 10^{10}$ & $1.74 \times 10^{11}$ & $1.83 \times 10^{10}$ & $2.30 \times 10^{11}$ \\
\hline
$0.14$ & $6.49\times 10^{10}$ & $1.96 \times 10^{11}$ & $2.05 \times 10^{10}$ & $2.59 \times 10^{11}$ \\
\hline
$0.145$ & $7.50 \times 10^{10}$ & $2.22 \times 10^{11}$ & $2.32 \times 10^{10}$ & $2.90 \times 10^{11}$ \\
\hline
$0.15$ & $8.76 \times 10^{10}$ & $2.54 \times 10^{11}$ & $2.64 \times 10^{10}$ & $3.30 \times 10^{11}$ \\
\hline
$0.155$ & $1.03 \times 10^{11}$ & $2.93 \times 10^{11}$ & $3.04 \times 10^{10}$ & $3.84 \times 10^{11}$ \\
\hline
$0.16$ & $1.22 \times 10^{11}$ & $3.41 \times 10^{11}$ & $3.52 \times 10^{10}$ & $4.48 \times 10^{11}$ \\
\hline
$0.165$ & $1.45 \times 10^{11}$ & $4.00 \times 10^{11}$ & $4.11 \times 10^{10}$ & $5.25 \times 10^{11}$ \\
\hline
$0.17$ & $1.75 \times 10^{11}$ & $4.74 \times 10^{11}$ & $4.84 \times 10^{10}$ & $6.20 \times 10^{11}$ \\
\hline
$0.175$ & $2.13 \times 10^{11}$ & $5.67\times 10^{11}$ & $5.77 \times 10^{10}$ & $7.41 \times 10^{11}$ \\
\hline
$0.18$ & $2.61 \times 10^{11}$ & $6.85\times 10^{11}$ & $6.94 \times 10^{10}$ & $8.96 \times 10^{11}$ \\
\hline
$0.185$ & $3.24 \times 10^{11}$ & $8.39\times 10^{11}$ & $8.44 \times 10^{10}$ & $1.09 \times 10^{12}$ \\
\hline
$0.19$ & $4.07 \times 10^{11}$ & $1.04 \times 10^{12}$ & $1.04 \times 10^{11}$ & $1.35 \times 10^{12}$ \\
\hline 
\end{tabular}
\caption{Values of $\lambda_{h \phi}/\sqrt{\lambda_{\phi}}$ and the corresponding values of $\lambda_{\phi}^{1/4} f_{a}$, for values of $\lambda_{\phi}^{1/4} f_{a}$ for which dark matter can be due to axions in the case of PQ symmetry breaking after inflation.}
\end{center}
\end{table}

\begin{figure}[h]
\label{Figure 2}
\begin{center}
\includegraphics[scale = 0.5]{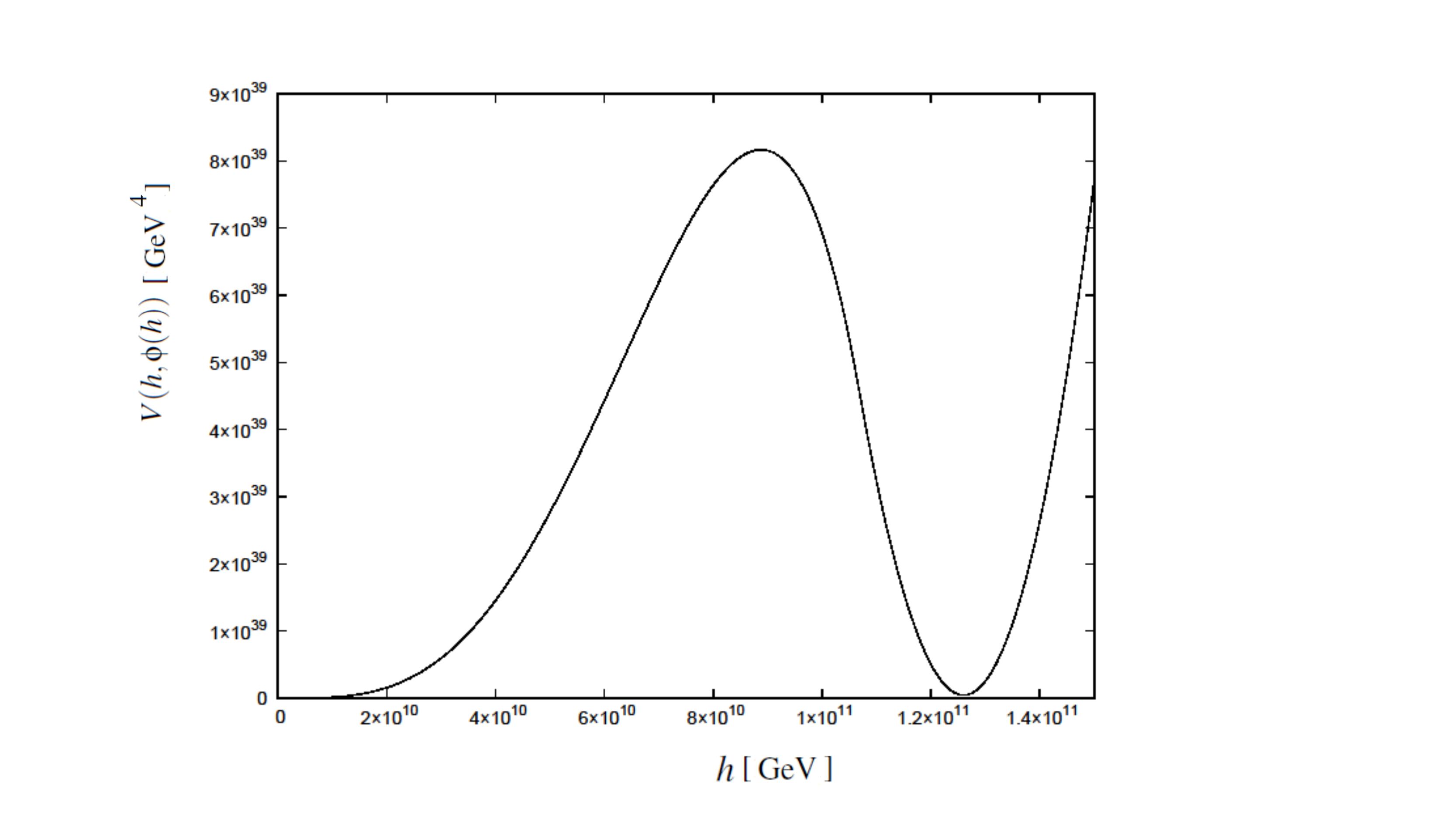}
\caption{The scalar potential with quasi-degenerate minima along the minimum trajectory when $\lambda_{h \phi}/\sqrt{\lambda_{\phi}}$ is at the lower bound for quasi-degenerate minima to exist, $\lambda_{h \phi}/\sqrt{\lambda_{\phi}} = 0.1$.}
\end{center}
\end{figure}

\begin{figure}[h]
\label{Figure 3}
\begin{center}
\includegraphics[scale = 0.5]{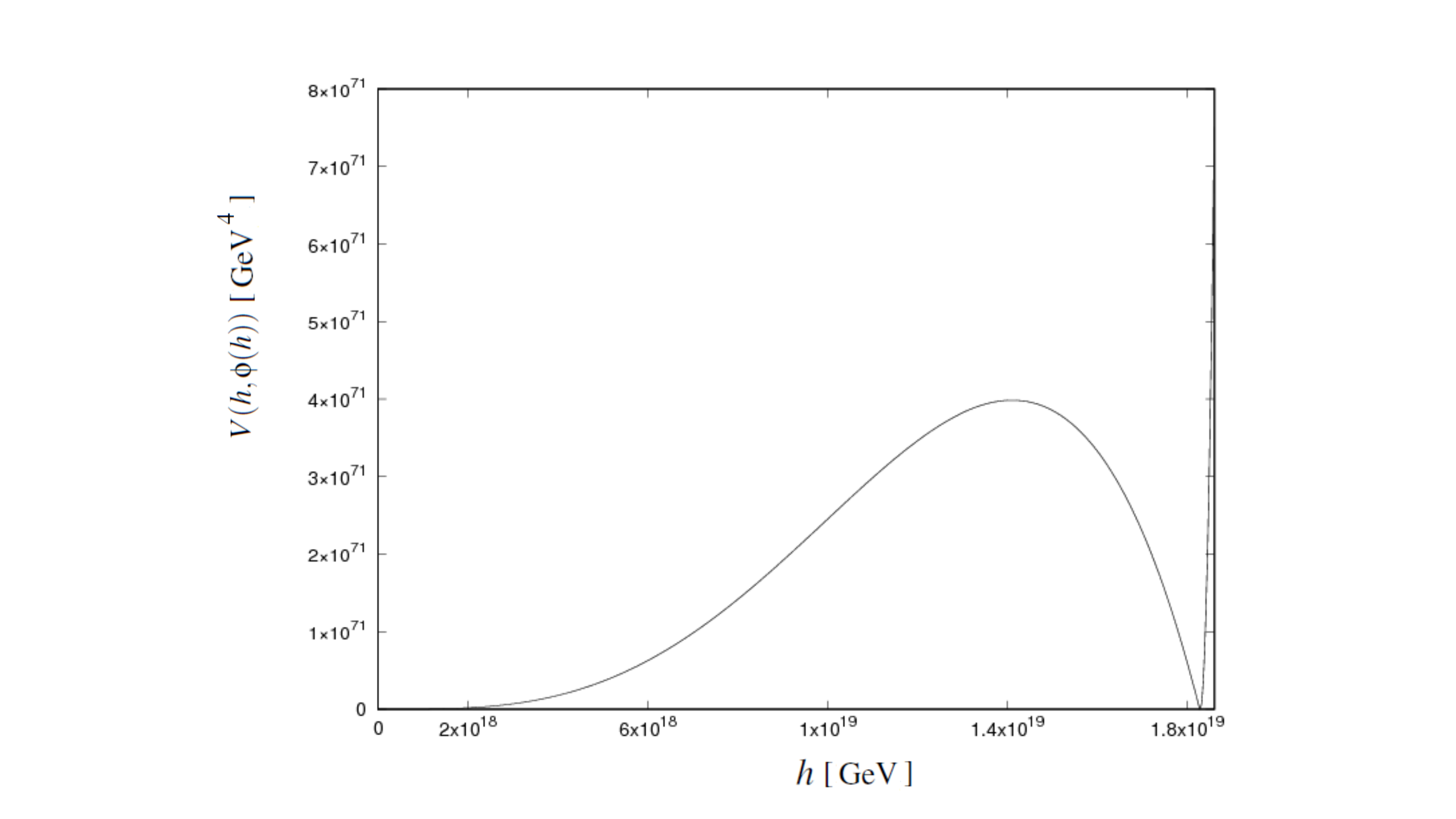}
\caption{The scalar potential with quasi-degenerate minima along the minimum trajectory when $\lambda_{h \phi}/\sqrt{\lambda_{\phi}}=0.297$, corresponding to the upper bound at which the potential minimum is close to the Planck scale.}
\end{center}
\end{figure}

\begin{figure}[h]
\label{Figure 4}
\begin{center}
\includegraphics[scale = 0.5]{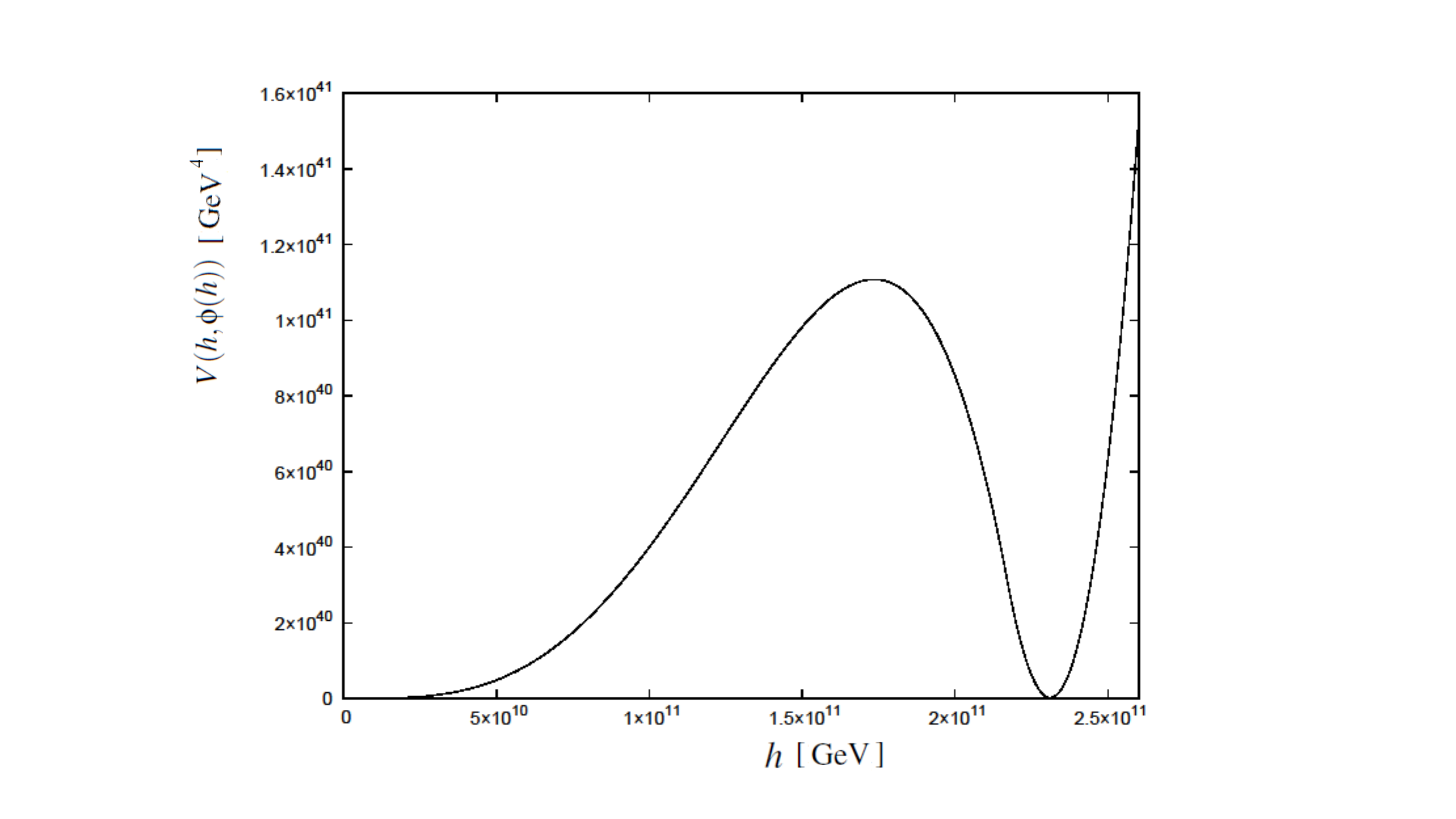}
\caption{The scalar potential with quasi-degenerate minima along the minimum trajectory for the case where $\lambda_{h \phi}/\sqrt{\lambda_{\phi}}=0.155$. This corresponds to $\lambda_{h \phi}^{\frac{1}{4}}f_{a}= 1.05 \times 10^{11} \GeV$, which is typical of the values for which axions can account for dark matter in the case where PQ symmetry is broken after inflation.}
\end{center}
\end{figure}

\begin{figure}[h]
\label{Figure 5}
\begin{center}
\includegraphics[scale = 0.5]{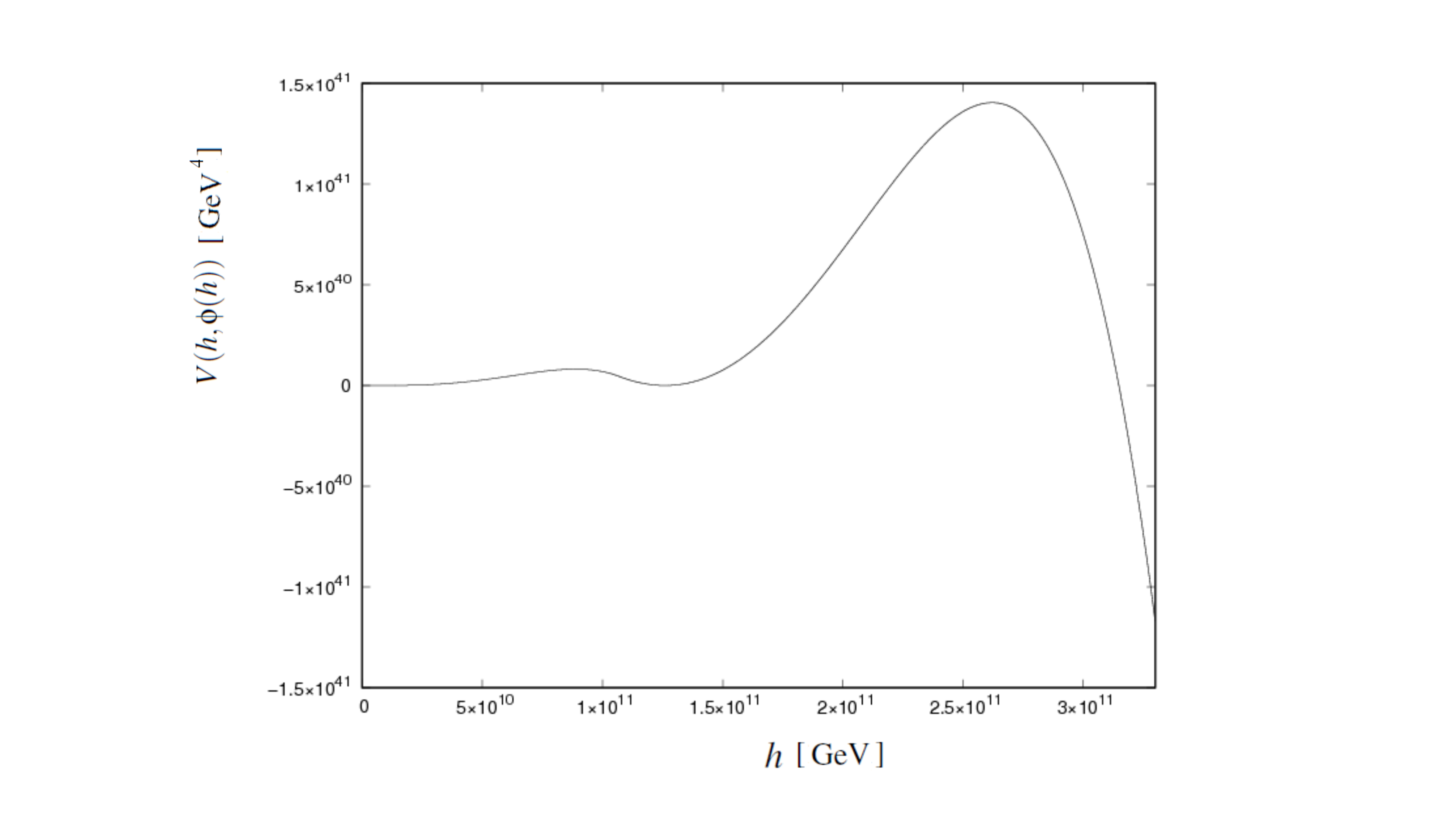}
\caption{The unbounded negative potential at large $h$, for the case $\lambda_{h \phi}/\sqrt{\lambda_{\phi}} = 0.1$.}
\end{center}
\end{figure}

\begin{figure}[h]
\label{Figure 6}
\begin{center}
\includegraphics[scale = 0.5]{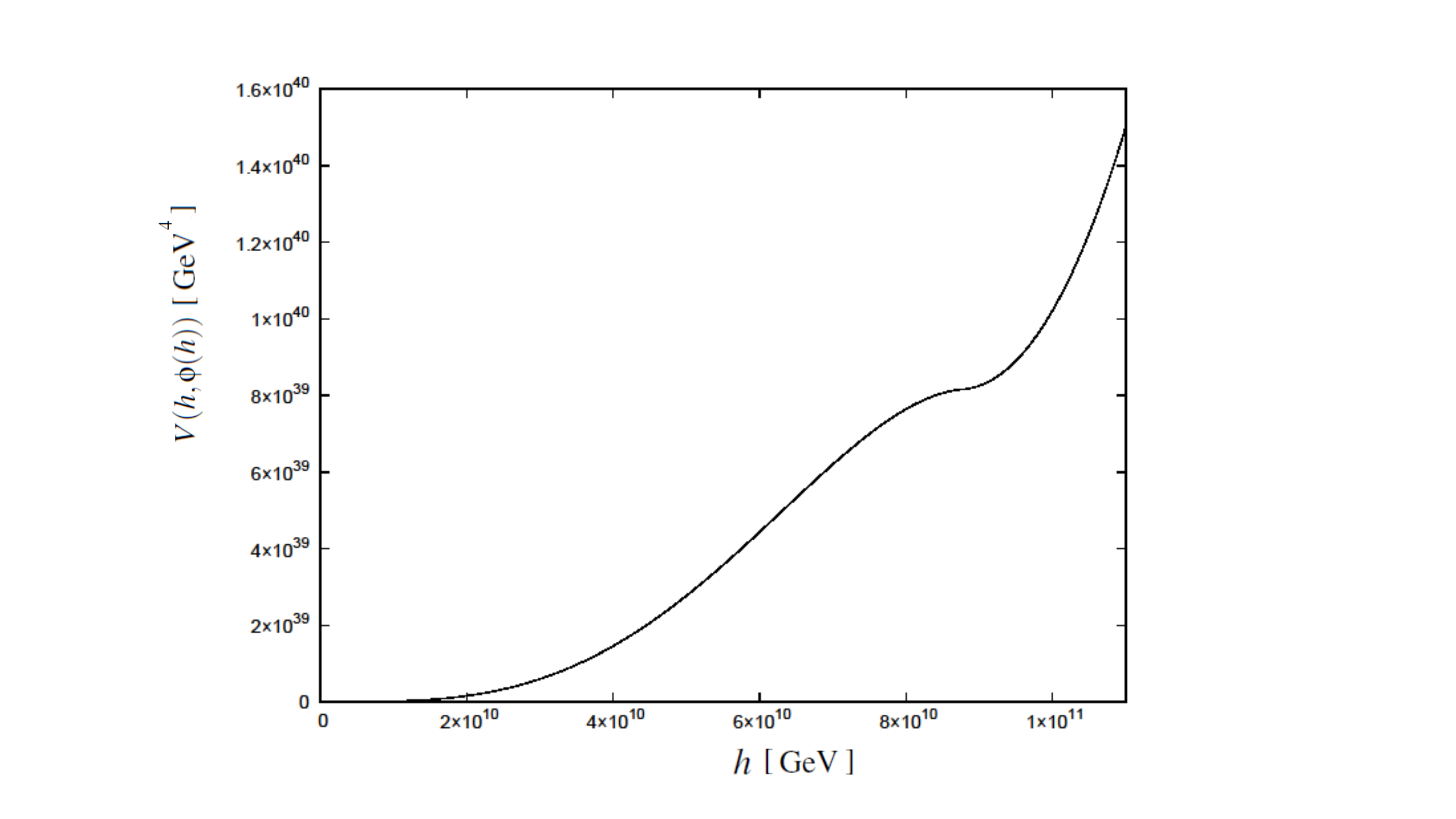}
\caption{An example of an inflection point in the potential at $\lambda_{h \phi}/\sqrt{\lambda_{\phi}} = 0.1$. Although we do not consider inflection points in this article, it is interesting to note that they can easily be found by reducing the value $\lambda^{1/4} f_{a}$ from the case of the degenerate potential.}
\end{center}
\end{figure}

In Figure 2 we show the potential along the minimum trajectory with a degenerate minimum at the lower limit of the $\lambda^{1/4}_{\phi} f_{a}$ range, corresponding to $\lambda_{h \phi}/\sqrt{\lambda_{\phi}} = 0.1$. In Figure 3 we show a degenerate minimum with $h_{min}$ close to the Planck scale, corresponding to $\lambda_{h \phi}/\sqrt{\lambda_{\phi}} = 0.297$. In Figure 4 we show the potential for a value of $\lambda_{\phi}^{1/4} f_{a}$ which is typical of the case where dark matter is entirely due to axions in the case of PQ symmetry breaking after inflation. 
In Figure 5 we show that the second minimum in the case with $\lambda_{h \phi}/\sqrt{\lambda_{\phi}} = 0.1$ is metastable, with an unbounded negative potential at large $h$.

In Figure 6 we show that it is also possible to obtain inflection point behaviour of the potential, by reducing $\lambda_{\phi}^{1/4} f_{a}$ for a given $\lambda_{h \phi}/\sqrt{\lambda_{\phi}}$ below the case of a degenerate potential. This may have applications to inflection-point inflation in the context of the KSVZ axion model, although we will not explore this possibility further here.

In order to determine the degree of tuning of $\lambda_{h \phi}/\sqrt{\lambda_{\phi}}$ required in order to explain the observed dark energy density, we have determined the value of the potential in the electroweak vacuum, $V_{SM}$, as a function of $\lambda_{h \phi}/\sqrt{\lambda_{\phi}}$ for a fixed value of $\lambda_{\phi}^{1/4} f_{a}$. We can express $V_{SM}$ as a Taylor expansion in the shift $\Delta (\lambda_{h \phi}/\sqrt{\lambda_{\phi}})$ from the value of $\lambda_{h \phi}/\sqrt{\lambda_{\phi}}$ for which $V_{SM} = 0$,
\be{de1} V_{SM} = k_{1} \,\Delta \left(\lambda_{h \phi}/\sqrt{\lambda_{\phi}}\right) + k_{2}\,\Delta \left(\lambda_{h \phi}/\sqrt{\lambda_{\phi}}\right)^{2}  + ... ~.\ee
In this we have defined the energy of the degenerate minimum to be zero, so that a change in $\lambda_{h \phi}/\sqrt{\lambda_{\phi}}$ from its degenerate value creates a change in $V_{SM}$. In general we find that the shift required in order to account for the observed dark energy density, $\rho_{vac} = 3.9 \times 10^{-47} \GeV^4$, is $\Delta (\lambda_{h \phi}/\sqrt{\lambda_{\phi}}) \, \sim \, \rho_{vac}/f_{a}^{4}$. For example, for the case in Table 2 corresponding to a degenerate vacuum with $\lambda_{h \phi}/\sqrt{\lambda_{\phi}} = 0.11$ and $\lambda_{\phi}^{1/4} f_{a} = 3.00 \times 10^{10} \GeV$, we find that $k_{1} = -5.6 \times 10^{42} \GeV^4$ and $k_{2} = 3.5 \times 10^{44} \GeV^{4}$. Therefore, in order to account for the observed dark energy density, we require a shift $\Delta \left(\lambda_{h \phi}/\sqrt{\lambda_{\phi}} \right) \approx -5 \times 10^{-90}$. This extremely small shift is a consequence of the large difference between the energy scale of the particle physics model, $f_{a}$,  and that of the observed dark energy density. Such tunings are likely to be a feature of any  explanation of the observed dark energy density based on a metastable particle physics vacuum. 

While the model requires fine-tuning, the naturalness of the tuning depends upon the underlying mechanism responsible for the near degeneracy of the vacuum states. The observed dark energy density is close to the present baryon and dark matter densities. This coincidence suggests that the dark energy density may be determined by a selection process acting on a landscape of metastable dark energy models with different vacuum energy splittings. In this case the extreme fine-tuning of the couplings  would be a natural outcome of the selection process.

\section{Axion Cosmology with a Quasi-Degenerate Potential}

If the PQ symmetry is broken during inflation such that the universe is in the electroweak vacuum $(h, \phi) = (v, f_{a})$ when observable scales exit the horizon ($N \lesssim 60$), and if axions account for dark matter, then there is a strong constraint from axion isocurvature perturbations. The isocurvature contribution to the total curvature plus entropy power spectrum, $\alpha$, due to axions is \cite{beltran} 
\be{i1} \alpha = \left( \frac{\Omega_{a}}{\Omega_{dm}}\right)^{2} \frac{8 M_{Pl}^{2} \epsilon }{f_{a}^{2}\theta_{a}^{2}}  ~,\ee 
where $M_{Pl} = (8 \pi G)^{-1/2} = 2.4 \times 10^{18} \GeV$, $\theta_{a}$ is the initial displacement of the axion field angle ($\theta_{a} = a/f_{a}$) from the minimum of its potential, and $\Omega_{a}$ and $\Omega_{dm}$ are the density in axions and the observed dark matter density respectively.
The present bound on isocurvature perturbations is $\alpha < \alpha_{lim} = 0.03$ \cite{planck}. Using $r = 16 \epsilon$ for the tensor-to-scalar ratio $r$, we obtain
\be{i2}  r \lesssim 2 \alpha_{lim} \left( 
\frac{\Omega_{dm}}{\Omega_{a}}\right)^{2} \frac{f_{a}^{2}  \theta_{a}^{2}}{M_{Pl}^{2} \epsilon}  ~.\ee
The axion dark matter density due to the misalignment mechanism is given by Eq.(2.8) of \cite{kawasaki}
\be{i3} \Omega_{a}^{mis}  = 3.8 \times 10^{-4}  \theta_{a}^2  \left( \frac{f_{a}}{10^{12} \GeV} \right)^{1.165}   ~.\ee
This assumes that the axion mass has the form given in Eq.(22)  of \cite{wantz} with $\Lambda_{QCD} = 400$ MeV and $n = 8.16$ \cite{qcd,ringwald}. Combining \eq{i2} and \eq{i3} gives an upper bound on $r$ from isocurvature perturbations, 
\be{i4} r \lesssim 7.2 \times 10^{-9} \left(\frac{\alpha_{lim}}{0.03}\right) \left( \frac{\Omega_{dm}}{\Omega_{a}}\right) \left( \frac{f_{a}}{10^{16} \GeV} \right)^{0.835}  ~.\ee
$r$ is related to the expansion rate during inflation at observable scales, $H_{inf}$, by 
\be{i5} H_{inf} = \pi M_{Pl} \left( \frac{{\cal P}_{R}\, r}{2}\right)^{1/2} = 7.9 \times 10^{13} \left(\frac{r}{0.1}\right)^{1/2} \GeV   ~,\ee
using ${\cal P}_{R}^{1/2} = 4.7 \times 10^{-5}$ for the observed curvature power. 
The isocurvature perturbation upper bound on $r$ then implies an upper bound on the rate of expansion during inflation, $H_{inf}$, given by 
\be{i6} H_{inf} \lesssim 2.1 \times 10^{10} \left(\frac{\alpha_{lim}}{0.03}\right)^{1/2} \left( \frac{\Omega_{dm}}{\Omega_{a}}\right)^{1/2} \left( \frac{f_{a}}{10^{16} \GeV} \right)^{0.418} \GeV  ~.\ee
Inflation models commonly have $H_{inf}$ larger than $10^{10} \GeV$ (i.e. $r$ larger than $10^{-8}$). In such models, in the case where there is a significant amount of axion dark matter (in particular, where axions account for all of the dark matter) and $f_{a}$ is small compared to $M_{Pl}$, the PQ symmetry must be restored and subsequently broken after inflation in order to satisfy the isocurvature constraint. 

In addition, and independently of the isocurvature constraint,  if the fields are at the quasi-degenerate minimum during inflation, then the electroweak and PQ symmetry must be restored after inflation to allow the fields to subsequently evolve to the electroweak vacuum. Therefore, if the fields are at either of the vacuum states during inflation (and if axions account for a significant amount of dark matter in the case of the electroweak vacuum), the PQ symmetry must be restored and subsequently broken after inflation.  

In the  case of PQ symmetry breaking after inflation\footnote{This assumes that finite-temperature symmetry breaking from $(h, \phi)  = (0,0)$ is towards the electroweak vacuum $(h, \phi)  = (v,f_{a})$, rather than the degenerate minimum $(h, \phi) \approx (h_{c}, 0)$. For small $\lambda_{h\phi}$ and $\lambda_{\phi}$ (as assumed in our analysis), the $T^2 \phi^2$ contribution to the finite-temperature potential will be small compared to the $T^2 h^2$ term. Therefore we expect that the symmetry will break along the $\phi$ direction towards the electroweak vacuum.}, the axion dark matter density is due to vacuum realignment and the decay of the cosmic strings and domain walls which form after PQ symmetry breaking and the QCD transition. The resulting dark matter density is related to $f_{a}$ by Eq.(4.4) of \cite{kawasaki} (using $n = 8.16$)
\be{c2}  \Omega_{a} h^2 \approx (1.6 \pm 0.4) \times 10^{-2} \left( \frac{f_{a}}{10^{10} \GeV} \right)^{1.165}  ~.\ee
Since $\lambda_{\phi}^{1/4} f_{a} \geq 2.39 \times 10^{10} \GeV$ is necessary in order for a degenerate minimum to exist, we find a lower bound on the amount of axion dark matter in the case where PQ symmetry is broken after inflation
\be{c3} \Omega_{a} \geq  (0.28-0.46) \lambda_{\phi}^{-0.291} \Omega_{dm} ~,\ee
where $\Omega_{dm} h^2 = 0.12$ is the observed dark matter density.
Thus if $\lambda_{\phi} \lesssim 1$ then at least $30\%$ of the dark matter must be due to axions\footnote{A similar result will be obtained in the case of PQ breaking during inflation in the case where the value of $\theta_{a}$ in our vacuum has a typical value, $\theta_{a} \sim \pi$, since in the case of PQ breaking after inflation we use the root mean square average of $\theta_{a}$ over domains after the PQ breaking transition, $\overline{\theta}_{a} \approx (\sqrt{2/3})\pi$ \cite{kawasaki}.}. This lower bound is a direct consequence of the assumption that dark energy is due to the energy of the electroweak vacuum relative to a second minimum at zero energy\footnote{Our axion dark matter bounds are based on the numerical simulation of axion strings and domain walls given in \cite{kawasaki}. We note that there have been very recent developments in axion simulations, which may modify the relation between $f_{a}$ and the amount of axion dark matter \cite{new1,new2}. In particular, \cite{new1} indicates a smaller number of axions from cosmic string decay, which would increase the value of $f_{a}$ necessary to account for a given fraction of dark matter.}.

If all of the dark matter is due to axions, then the value of $f_{a}$ is fixed by \eq{c2} to be in the range $4.7 \times 10^{10} \GeV$ to $7.2 \times 10^{10} \GeV$. The potential is determined by the value of $\lambda_{h \phi}/\sqrt{\lambda_{\phi}}$, which is fixed (via the degenerate minimum constraint) by the value of $\lambda_{\phi}^{1/4} f_{a}$ via Table 1. Therefore, up to a weak dependence on $\lambda_{\phi}$ (and uncertainties in the axion dark matter density from cosmic string and domain wall decay),  the form of the potential is fixed. This will allow the cosmology of the model to be quantitatively studied. For $\lambda_{\phi}$ in the dimensionally natural range 0.01-1, the uncertainty in $ \lambda_{\phi}^{1/4} f_{a}$ due to the unknown value of $\lambda_{\phi}$ corresponds to a factor between 0.32 and 1. 

\section{Conclusions and Discussion} 

We have shown that dark energy in the KSVZ axion model can be understood as being due the energy density of a metastable electroweak vacuum relative to a second  minimum of the potential which is at zero energy density.  The resulting model is therefore able to account for both the dark matter and dark energy densities as well as to provide a solution to the strong CP problem. The requirement of a second almost degenerate minimum of the scalar potential imposes a non-trivial constraint on the parameters of the model. In particular, it implies a lower bound on the axion decay constant for a given PQ scalar self-coupling via a lower bound on $\lambda_{\phi}^{1/4} f_{a}$. In the case where PQ symmetry is broken after inflation, this implies a lower bound on the amount of axion dark matter of about 30$\%$ of the total dark matter density if $\lambda_{\phi} \lesssim 1$. In addition, if axions constitute all of the observed dark matter, then the value of $f_{a}$ is fixed (up to uncertainties in the production of axions from cosmic string and domain wall decay), which fixes the form of the potential up to a weak dependence on $\lambda_{\phi}^{1/4}$. This will allow the cosmology of the KSVZ model with axion dark matter and metastable dark energy to be quantitatively studied.    

   The energy density of the lowest energy metastable or stable minimum being equal to zero can arise, for example, in theoretical approaches which explain the cancellation of the vacuum energy density via spacetime averaging, such as vacuum sequestering, and in models with an energy parity.
 Both of these approaches (more generally, any approach that effectively subtracts off the total energy of one of the vacuum states) will ensure that there is no cosmological constant term, as any constant term in the potential is cancelled. Importantly, this ensures that the large quantum contribution to the vacuum energy density is also cancelled. Dark energy must then be explained either by adding a new quintessence component to the energy density, or by a vacuum energy density due to the difference in energy between a pair of vacuum states. We have shown that the KSVZ axion model can provide the necessary pair of vacuum states. The KSVZ axion model with quasi-degenerate minima can therefore serve as a minimal model for dark energy, in the sense that it does not require any additional component to account for dark energy. In this context, the question of the smallness of the observed dark energy density, which requires the couplings to be tuned to a precision of the order of $\rho_{vac}/f_{a}^4$, becomes the question of the underlying physics responsible for the near-degeneracy of the vacuum states.

 In this first study we have not included the quantum corrections to the scalar potential due to the PQ scalar couplings $\lambda_{h \phi}$ and $\lambda_{\phi}$. This will be a good approximation if the classical modification of the potential due to the $\phi$ field and its mixing with the electroweak doublet Higgs field are the dominant effects. A complete analysis of the effective potential 
requires a multiscale RG analysis of the two-field scalar potential in order to keep the CW correction to the potential small for all values of $h$ and $\phi$. 
In addition, our analysis has been at next-to-leading order (NLO) in the SM effective potential. 
Next-to-next-to-leading order (NNLO) corrections are known to significantly increase the stability of the SM potential \cite{degrassi}. 
We will return to the full analysis of the KSVZ effective potential, using multiscale RG methods and including NNLO corrections, in future work.

   It has recently been suggested that, in generalizations of the SM Higgs potential which have a second minimum due to the scalar threshold effect, primordial black holes can form via quantum fluctuations \cite{pbh}. The KSVZ axion model with a degenerate potential, PQ symmetry breaking after inflation and axion dark matter, would therefore provide a well-defined potential of the required form. We will consider this in a future analysis. 

More generally, it would be interesting to explore the possible cosmological evolution of the model in different inflation scenarios. In particular, the non-minimally coupled PQ scalar inflation model of \cite{marsh} would be interesting in this context. Combined with the quasi-degenerate vacuum explanation of the dark energy density, this would provide a minimal model for dark matter, dark energy and inflation, as well as a solution to the strong CP problem\footnote{An alternative approach, which unifies axion dark matter with inflation, neutrino masses and baryogenesis, was proposed in \cite{smash,smash2}.}.

\section*{Acknowledgements} The work of ALS was supported by STFC. The work of JM was partially supported by STFC via the Lancaster-Manchester-Sheffield Consortium for Fundamental Physics.

\end{document}